\documentclass[twocolumn]{aastex631}
\usepackage{amsmath,amssymb,bm}
 
\begin{document}

\title{Enhanced Magnetic Field Amplification by Ion-Beam Weibel Instability in Weakly Magnetized Astrophysical Shocks}

\author[0000-0002-1876-5779]{Taiki Jikei}
\affiliation{Department of Earth and Planetary Science, The University of Tokyo, 7-3-1 Hongo, Bunkyo-ku, Tokyo, 113-0033, Japan}

\author[0000-0002-2140-6961]{Takanobu Amano}
\affiliation{Department of Earth and Planetary Science, The University of Tokyo, 7-3-1 Hongo, Bunkyo-ku, Tokyo, 113-0033, Japan}

\author[0000-0002-1484-7056]{Yosuke Matsumoto}
\affiliation{Institute for Advanced Academic Research, Chiba University, 1-33 Yayoi-cho, Inage-ku, Chiba, 263-8522, Japan}

\begin{abstract}
We examine the evolution of ion-beam Weibel instability at strong collisionless shocks in weakly magnetized media. 
We find that a finite background magnetic field substantially affects both linear and nonlinear phases of the instability, depending on whether the background electrons behave magnetized or not. 
Particle-in-cell simulations for magnetized electrons identify a dynamo-like mechanism of magnetic field amplification, which eventually leads to spontaneous magnetic reconnection. 
We conclude that this scenario is applicable to typical young supernova remnant shocks.
\end{abstract}

\section{Introduction} \label{sec:intro}
Collisionless shocks have been extensively studied over the decades in the context of space physics, astrophysics, as well as laboratory astrophysics \citep{Balogh2013}. 
One of the prime motivations is to understand the acceleration of high-energy cosmic rays at distant astrophysical shocks, such as in supernova remnants (SNRs), active galactic nuclei (AGN), and gamma-ray bursts (GRBs). 
A wide variety of kinetic plasma instabilities at various scales may be activated and play a role in particle acceleration depending on the shock parameters \citep{Wu1984a}. 
In particular, microscopic instabilities driven by the shock-reflected ion beam in the shock transition region have been thought of as the prime agent to resolve the long-standing electron injection problem \citep{McClements2001,Hoshino2002,Amano2007,Riquelme2011}. 
Among all the candidates, Weibel instability \citep{Weibel_1959} has attracted great attention as the dominant mode at a very high Mach number regime \citep{Medvedev_1999,Silva_2003,Hededal_2004,Frederiksen_2004,Medvedev_2004,Kato_2007,Kato_2008,Spitkovsky_2008,Kato_2010,Fiuza2012,Matsumoto2017,Takamoto_2018,Fiuza_2020,Fox_2013,Huntington_2015}.

Earlier studies of Weibel instability in the context of collisionless shocks focused on relativistic shocks in unmagnetized plasmas \citep{Medvedev_1999,Silva_2003,Hededal_2004,Frederiksen_2004,Medvedev_2004}. 
Particle-In-Cell (PIC) simulations both in two dimensions (2D) and three dimensions (3D) demonstrated that current filaments generated in the early phase of the instability subsequently merge with each other in the long-term evolution to convert a fraction of the shock kinetic energy into the magnetic energy progressively at larger and larger scales \citep{Kato_2005,Ruyer2015,Ruyer2016}. 
\citet{Ruyer_2018} recently also showed that the current filaments might eventually break up by the kink instability in a realistic, fully 3D system. 
On the other hand, 2D PIC simulations of high-Mach-number non-relativistic astrophysical shocks with a weak but finite background magnetic field showed a somewhat different consequence \citep{Matsumoto_2015,Bohdan2017,Bohdan_2020_III,Bohdan_2021}. 
While the dominant instability is still reasonably understood as an ion-beam Weibel instability, the magnetic field component contained within the 2D simulation plane is amplified to even larger than the out-of-plane component, which should be the dominant component as predicted by the conventional Weibel instability in an unmagnetized plasma \citep{Bohdan_2021}. 
The amplification of the in-plane magnetic field spontaneously produces coherent current sheets, which eventually break up via magnetic reconnection within the shock transition region. 
How the two different types of nonlinear evolution of Weibel instability in the relativistic and non-relativistic shocks are discriminated has not been understood. 
In this paper, we demonstrate that the linear and nonlinear evolution of the ion-beam Weibel instability at high-Mach-number magnetized collisionless shocks is regulated crucially by the magnetization of the background electron component.

\newpage
\section{Theory} \label{sec:theory}
\subsection{Linear Growth Rate} \label{subsec:linear}
To mimic the transition region of a magnetized perpendicular shock, we consider a homogeneous system consisting of three particle populations: the background electrons, the reflected ions, and the incoming ions, all represented by the isotropic drifting Maxwellian distribution.
We work in the rest frame of electrons, which approximately corresponds to the upstream plasma frame.
The charge neutrality and zero net current conditions indicate that the density and bulk velocity are given respectively by $Nn_0, (1-N)V_{\mathrm{sh}}$ for the reflected ions and $(1-N)n_0, -NV_{\mathrm{sh}}$ for the incoming ions, where $n_0$ is the electron density, $N$ is the fraction of the reflected ion component, and $V_{\mathrm{sh}}$ is the shock propagation speed in the upstream frame, respectively.
The background magnetic field with the strength of $B_0$ points perpendicular to the ion beam.

Strictly speaking, this setup is not in equilibrium as the ion velocity distribution is asymmetric in the plane perpendicular to the background magnetic field. Therefore, we will focus only on the fast time scale phenomena occurring over the time scale $T\lesssim \Omega_i^{-1}$, where $\Omega_s$ is the cyclotron frequency of particle species $s$. 
Indeed, it has been shown that the reflected-ion-driven mode becomes Weibel-like when the maximum growth rate satisfies the condition $\Gamma_{\mathrm{max}}/\Omega_i \gg 1$, which requires sufficiently high Mach number shocks relevant to, e.g., SNRs \citep{Nishigai_2021}. 
In other words, we consider only this specific parameter regime where ions behave unmagnetized. 
On the other hand, the background electrons may or may not behave magnetized depending on the parameter $\Gamma_{\mathrm{max}}/\Omega_e$. 
Henceforth, the magnetized electron refers to the condition $\Gamma_{\mathrm{max}}/\Omega_e \ll 1$, whereas the unmagnetized electron refers to the opposite $\Gamma_{\mathrm{max}}/\Omega_e \gtrsim 1$. 
As we shall see later, the electron magnetization will be determined by the Alfv\'{e}n Mach number of the shock.

For simplicity, we use the cold plasma approximation \citep{Stix_1992}.
Note that we consider a finite thermal spread in the simulations. 
However, we confirmed that the linear growth rates of the dominant modes (near the maximum growth rate) were consistent with the cold plasma approximation. 
We assume the ions are unmagnetized $(\Omega_i/\omega\sim0)$, and the electrons are magnetized.
The background magnetic field is set to the $z$-direction $\bm{B}_0=B_0\hat{\bm{e}}_z$, and the wavenumber vector is defined as $\bm{k}=k(0,\sin\theta,\cos\theta)^{\mathsf{T}}$.
We first calculate the conductivity of the ion species $s$, which has the unperturbed density of $N_sn_0$ and unperturbed velocity of $\bm{V}_{0s}=V_{0s}\hat{\bm{e}}_x$.
The linearized continuity equation and the equation of motion read
\begin{align}
-i\omega\tilde{n}_s+ikN_sn_0(\sin\theta \tilde{u}_{sy}+\cos\theta \tilde{u}_{sz})=0, \\
-i\omega \tilde{u}_{sx} -\frac{e}{m_i}\tilde{E}_x=0, \\
-i\omega \tilde{u}_{sy} -\frac{e}{m_i}(\tilde{E}_y-V_{0s}\tilde{B}_z/c)=0, \\
-i\omega \tilde{u}_{sz} -\frac{e}{m_i}(\tilde{E}_z+V_{0s}\tilde{B}_y/c)=0.
\end{align}
Note that, since we assume that the ions are unmagnetized, we ignore the Lorentz force in the ion equation $\tilde{\bm{v}}\times\bm{B}_0/c$ although this is also a first-order term in the strict sense.
The conductivity tensor $\bm{\sigma}_s$ which is defined by $\bm{j}_s=eN_sn_{0}\tilde{\bm{v}}_s+e\tilde{n}_s\bm{V}_{0s}=\bm{\sigma}_s\cdot\bm{\tilde{E}}$ reads 
\begin{equation}
\begin{split}
&\bm{\sigma}_s[N_s,V_{0s}]=\\
&\frac{N_sn_0e^2}{m_i}\frac{i}{\omega}
\begin{bmatrix}
1+V_{0s}^2\left(\dfrac{k}{\omega}\right)^2 & V_{0s}\dfrac{k}{\omega}\cos\theta & V_{0s}\dfrac{k}{\omega}\sin\theta \\
V_{0s}\dfrac{k} {\omega}\cos\theta & 1 & 0 \\
V_{0s}\dfrac{k}{\omega}\sin\theta & 0 & 1
\end{bmatrix}.
\end{split}
\end{equation}
The conductivity tensor of the incoming and reflected ions can be obtained by replacing $N_s$ with $N$ and $(1-N)$, $V_{0s}$ with $(1-N)V_{\mathrm{sh}}$ and $-NV_{\mathrm{sh}}$ respectively.
We can calculate the total conductivity of the ions by adding the two tensors, which reads
\begin{equation}
\begin{split}
\bm{\sigma}_i&=\bm{\sigma}_s[N,(1-N)V_{\mathrm{sh}}]+\bm{\sigma}_s[1-N,-NV_{\mathrm{sh}}]\\
&=\frac{n_0e^2}{m_i}\frac{i}{\omega}
\begin{bmatrix}
1+[N(1-N)V_{\mathrm{sh}}]^2\left(\dfrac{k}{\omega}\right)^2 & 0 & 0 \\
0 & 1 & 0 \\
0 & 0 & 1
\end{bmatrix}.
\end{split}
\end{equation}
Note that the off-diagonal components cancel out after taking the sum.
The conductivity of cold magnetized electrons is well-known \citep{Stix_1992}: 
\begin{equation}
\bm{\sigma}_e=\frac{n_0e^2}{m_e}\frac{i}{\omega}
\begin{bmatrix}
\frac{\omega^2}{\omega^2-\Omega^2_e} & \frac{i\omega \Omega_e}{\omega^2-\Omega^2_e} & 0 \\
-\frac{i\omega \Omega_e}{\omega^2-\Omega^2_e} & \frac{\omega^2}{\omega^2-\Omega^2_e} & 0 \\
0 & 0 & 1
\end{bmatrix}.
\end{equation}
The sum of conductivity tensor $\bm{\sigma}=\bm{\sigma}_i+\bm{\sigma}_e$ describes the total response of this system.
By using this conductivity tensor and Maxwell's equations, we can calculate the dispersion tensor $\bm{D}$, which is defined by $\bm{D}\cdot\tilde{\bm{E}}=\bm{0}$.
\begin{widetext}
\begin{equation} \label{dispersiontensor}
\begin{split}
\bm{D}&=
\begin{bmatrix}
(k\lambda_e)^2\cos^2\theta+\frac{m_e}{m_i}+\frac{\omega^2}{\omega^2-\Omega^2_e} & i\frac{\omega\Omega_e}{\omega^2-\Omega^2_e} & -(k\lambda_e)^2\sin\theta\cos\theta \\
-i\frac{\omega\Omega_e}{\omega^2-\Omega^2_e} & (k\lambda_e)^2\cos^2\theta+\frac{m_e}{m_i}+\frac{\omega^2}{\omega^2-\Omega^2_e} & 0 \\
-(k\lambda_e)^2\sin\theta\cos\theta & 0 & (k\lambda_e)^2\sin^2\theta+\frac{m_i+m_e}{m_i}
\end{bmatrix}\\
&+
\begin{bmatrix}
N(1-N)(k\lambda_e)^2\left(\frac{V_{\mathrm{sh}}}{c}\right)^2\left(\frac{\omega_{pi}}{\omega}\right)^2&0&0\\
0&0&0\\
0&0&0
\end{bmatrix}.
\end{split}
\end{equation}
\end{widetext}
Where $\omega_{ps}=\sqrt{4\pi n_0 e^2/m_s}$ and $\lambda_s=c/\omega_{ps}$ are plasma frequency and inertial length of species $s$, respectively.
We assume $1\ll k^2c^2/\omega^2$, which corresponds to ignoring the displacement current.
We can obtain the analytic solution of $\mathrm{det}(\bm{D})=0$ by converting to a quadratic equation with respect to $X=\omega^2/(\omega^2-\Omega_e^2)$.
Note that $\omega$ that appears in different forms can be rewritten as $\omega=\Omega_e\sqrt{X/(1-X)}$.

Let us compare the magnetized and unmagnetized electron cases and discuss the effect of the background magnetic field.
We assume $m_e/m_i\ll1$ and consider parallel propagation $\theta=0$ to compare the magnetized and unmagnetized cases.
In the weak background magnetic field limit $(\Omega_e\to0)$, we obtain the growth rate of the unmagnetized Weibel instability
\begin{equation} \label{eq:unmag}
\frac{\omega}{\omega_{pi}}=i\sqrt{N(1-N)}\frac{V_{\mathrm{sh}}}{c}
\left( 1+\frac{1}{(k\lambda_e)^2} \right)^{-1/2}.
\end{equation}
Note that the factor $\left( 1+1/(k\lambda_e)^2 \right)^{-1/2}$ represents the electron screening effect, which will be discussed in detail in the next paragraph.
On the other hand, if we assume a background magnetic field strong enough to magnetize electrons in the sense $|\omega/\Omega_e|\ll1$ (but $|\omega/\Omega_i|\gg1$), we obtain the following dispersion relation.
\begin{equation}
\frac{\omega}{\omega_{pi}}=i\sqrt{N(1-N)}\frac{V_{\mathrm{sh}}}{c}.
\end{equation}
Note that the electron screening factor has disappeared.

Note that \citet{Grassi2017} investigated the effect of finite background magnetic field in a different setup.
They considered beam-aligned background magnetic field for electron beam.
The growth rate decreases, but the saturation stage is only weakly affected.
Their results do not contradict ours because the beam species (energy source) and the magnetic field orientation differ.
Studying the dependence on the strength of the magnetic field in a wider range (from a weak magnetic field that only magnetizes the electrons to a strong magnetic field that magnetizes the ions) and magnetic field orientation, including oblique angles, may be an important topic for future works.

\begin{figure}[htbp]
\centering
\includegraphics[width=\linewidth]{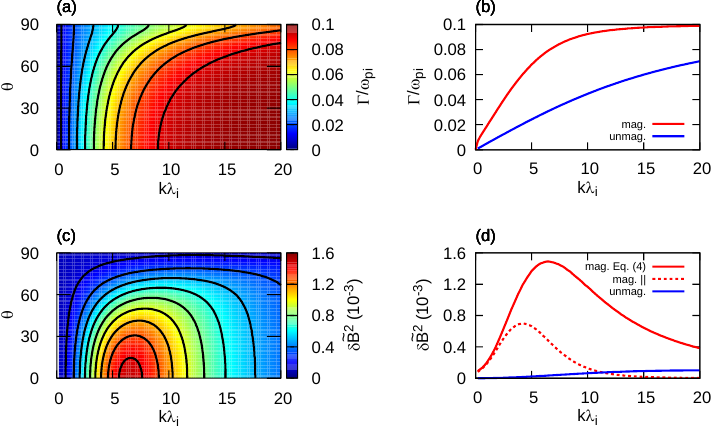}
\caption{\label{fig1} Growth rate and saturation level based on the trapping condition. (a) Growth rate $\Gamma/\omega_{pi}$ as a function of $k$ and $\theta$. (b) Growth rate at parallel propagation $\theta=0$. The red and blue lines correspond to the magnetized and unmagnetized electron models, respectively. (c) Saturation level $\delta \tilde{B}^2$ as a function of $k$ and $\theta$ evaluated based on the trapping condition. (d) Saturation level at parallel propagation $\theta =0$. The red and blue solid lines correspond to the magnetized and unmagnetized electron models. The red dotted line corresponds to the beam-aligned component for the magnetized electron model evaluated using the linear eigenvector.}
\end{figure}

The results of linear analysis obtained with parameters $m_i/m_e=400, N=0.2, V_{\mathrm{sh}}=0.25c$ for the magnetized ($\Omega_e/\omega_{pe}=0.05$) and the unmagnetized ($\Omega_e/\omega_{pe}=0$) electron models are shown in Fig. \ref{fig1}.
Fig. \ref{fig1} (a) shows the growth rate $(\Gamma=\mathrm{Im}[\omega])$ as a function of wavenumber $k$ and propagation angle $\theta$ with respect to the background magnetic field.
Fig. \ref{fig1} (b) compares the growth rate at the parallel propagation $(\theta=0)$ between the two cases.
We see that a finite background magnetic field introduces propagation angle dependence and increases the growth rate at long wavelengths ($k\lambda_i\lesssim10$ where $\lambda_i$ is the ion inertial length).
The shift in wavelength and the increase in the growth rate may be understood by the difference in the electron response.
When the electrons are unmagnetized, the inductive electric field $\delta \bm{E}$ accelerates the electrons, which then tend to screen out the ion current. 
Therefore, a larger growth rate should be obtained at a wavelength smaller than the electron inertial length beyond which the electron screening becomes inefficient \citep{Achterberg_2007,Ruyer2016}.
On the other hand, if the electrons are magnetized, the inductive electric field is no longer able to accelerate them parallel to it but induces the $\delta\bm{E}\times\bm{B}_0$ drift instead \citep{Nishigai_2021}.
The screening effect then becomes weak, especially at longer wavelengths, which results in larger growth rates.

\subsection{Saturation by Trapping}
It is known that the linear growth of Weibel instability ceases when the trapping condition
$\Gamma \sim \omega_{\mathrm{B}}=\left[k V_{\mathrm{sh}} \Omega_i (\delta B(k)/B_0)\right]^{1/2}$  is satisfied, where $\omega_{\mathrm{B}}$ is the bounce frequency which corresponds to the angular frequency of a particle trapped in the perturbed magnetic field $\delta B(k)$ \citep{Lutomirski_1966,Davidson_1972}.
The trapping condition may be rewritten as follows.
\begin{equation}
\delta \tilde{B}^2(k,\theta)=N^{-1}(1-N)^{-1}\left(\frac{\Gamma}{\omega_{pi}}\right)^4(k\lambda_i)^{-2}\left(\frac{V_{\mathrm{sh}}}{c}\right)^{-4}.
\label{eq:saturation_level}
\end{equation}
Note that the growth rate $\Gamma$ also contains dependence on $N$ and $V_{\mathrm{sh}}$.
The final form of saturation level is not sensitive to these parameters after canceling out.
Throughout this study, the magnetic field given in units of $B_{\mathrm{kin}}=\sqrt{4\pi N(1-N)n_0m_iV_{\mathrm{sh}}^2}$ will be denoted by a tilde, i.e., $\delta \tilde{B} \equiv \delta B / B_{\mathrm{kin}}$. 
Therefore, $\delta \tilde{B}^2$ gives the energy conversion rate from the initial ion kinetic energy to the magnetic energy.
Fig. \ref{fig1} (c) shows $\delta \tilde{B}^2$ in the $k-\theta$ plane.
The dominance of near parallel propagation and the shift in peak wavenumber from the electron inertial scale ($k\lambda_i\sim20$ with a mass ratio of $m_i/m_e=400$) to a longer wavelength $(k\lambda_i\sim6)$ is more obvious in this quantity.
This is explained by the higher growth rate and the scaling of the bounce frequency $\omega_{B}\propto k^{1/2}$.
Fig \ref{fig1} (d) compares the saturation level estimate at the parallel propagation $\delta\tilde{B}^2(k,\theta = 0)$.
The primary components $\delta \tilde{B}^2$ given by Eq. (\ref{eq:saturation_level}) are shown with the solid lines for the two cases.
The red dotted line is the normalized beam-aligned component energy at the saturation (see App. \ref{subsec:polarization} for the exact definition).
We see that the maximum saturation level is an order of magnitude larger for the magnetized case.
For $k \lambda_i \gtrsim 10$, the beam-aligned component is much smaller than the primary component, which implies that the background magnetic field affects the growth rate but not the polarization at a short wavelength.
At around the peak saturation level, the beam-aligned component is responsible for $\sim 20{\rm -}30\%$ of the magnetic energy.
Note that the beam-aligned component has comparable amplitudes at long wavelengths $k\lambda_i\lesssim4$.
However, we think that the small growth rate in this region will invalidate the assumption of unmagnetized ions, and it is not relevant to our application. We then conclude that the beam-aligned component appears finite in the presence of the magnetized electrons but remains subdominant in the linear regime.

Although the above estimate gives only a small saturation level ($\delta \tilde{B}^2 \sim 10^{-3}$), we should keep in mind that the trapping condition merely provides a rough estimate for quenching the linear growth. 
The nonlinear evolution of Weibel instability in an unmagnetized plasma shows further amplification of the magnetic field involving the merging of current filaments, which is triggered by the seed fluctuations generated in the early phase \citep{Ruyer2015}. 
Therefore, the quantitative difference in the linear property and the trapping-limited saturation level arising from the magnetized electron response motivates further investigation of the fully nonlinear evolution.

\section{PIC simulation} \label{sec:simulation}
To investigate the nonlinear evolution of the system, we performed 2D PIC simulations \citep{Matsumoto_2015} with the same parameters used for the linear analysis, except that we used $\Omega_e/\omega_{pe}=0.005$ instead of 0 for the unmagnetized electron case. 
We have confirmed that the magnetized case $\Omega_e/\omega_{pe}=0.05$ satisfies $\Gamma_{\mathrm{max}}/\Omega_e \ll 1$, whereas $\Omega_e/\omega_{pe}=0.005$ gives $\Gamma_{\mathrm{max}}/\Omega_e \sim 1$. 
The latter is thus suitably referred to as the unmagnetized electron case, even though the electron cyclotron frequency is non-zero.
The thermal velocity of each components were $v_{\mathrm{th},i}=0.00125c, v_{\mathrm{th},e}=0.1c$.
These parameters correspond to Alfv\'{e}n Mach numbers of $M_{\mathrm{A}}=V_{\mathrm{sh}}(c\Omega_i/\omega_{pi})^{-1}=1000, 100$ for the unmagnetized and magnetized electrons, respectively.
The sound Mach number $M_{\mathrm{s}}=V_{\mathrm{sh}}/v_{\mathrm{th},i}=200$ is the same for both cases.
Note that we used the same thermal velocity for both reflected and incoming ions for simplicity.
We used two different configurations for the beam direction relative to the simulation plane:
One is the out-of-plane beam, i.e., the simulation is in the $y-z$ plane, and the other is the in-plane beam, i.e., the simulation is in the $x-z$ plane.
We used a fixed grid size equal to the  Debye length $(\Delta x=\lambda_D=v_{\mathrm{th},e}/\omega_{pe})$.
The time step was determined by the condition $c\Delta t/\Delta x=1$ (or $\omega_{pe} \Delta t = 0.1$), which is ideal for suppressing the numerical Cherenkov instability \citep{Ikeya2015}.
Note that we use a semi-implicit Maxwell solver in our code \citep{Hoshino2013}.
The number of particles per cell was 32 for both ions (sum of reflected and incoming ions) and electrons.
The simulation box was a square with $4032 \times 4032$ grid points, corresponding to a side length of $20.16\lambda_i$.

\newpage
\subsection{Out-of-Plane Beam} \label{subsec:outofplane}
\begin{figure}[htbp]
\centering
\includegraphics[width=\linewidth]{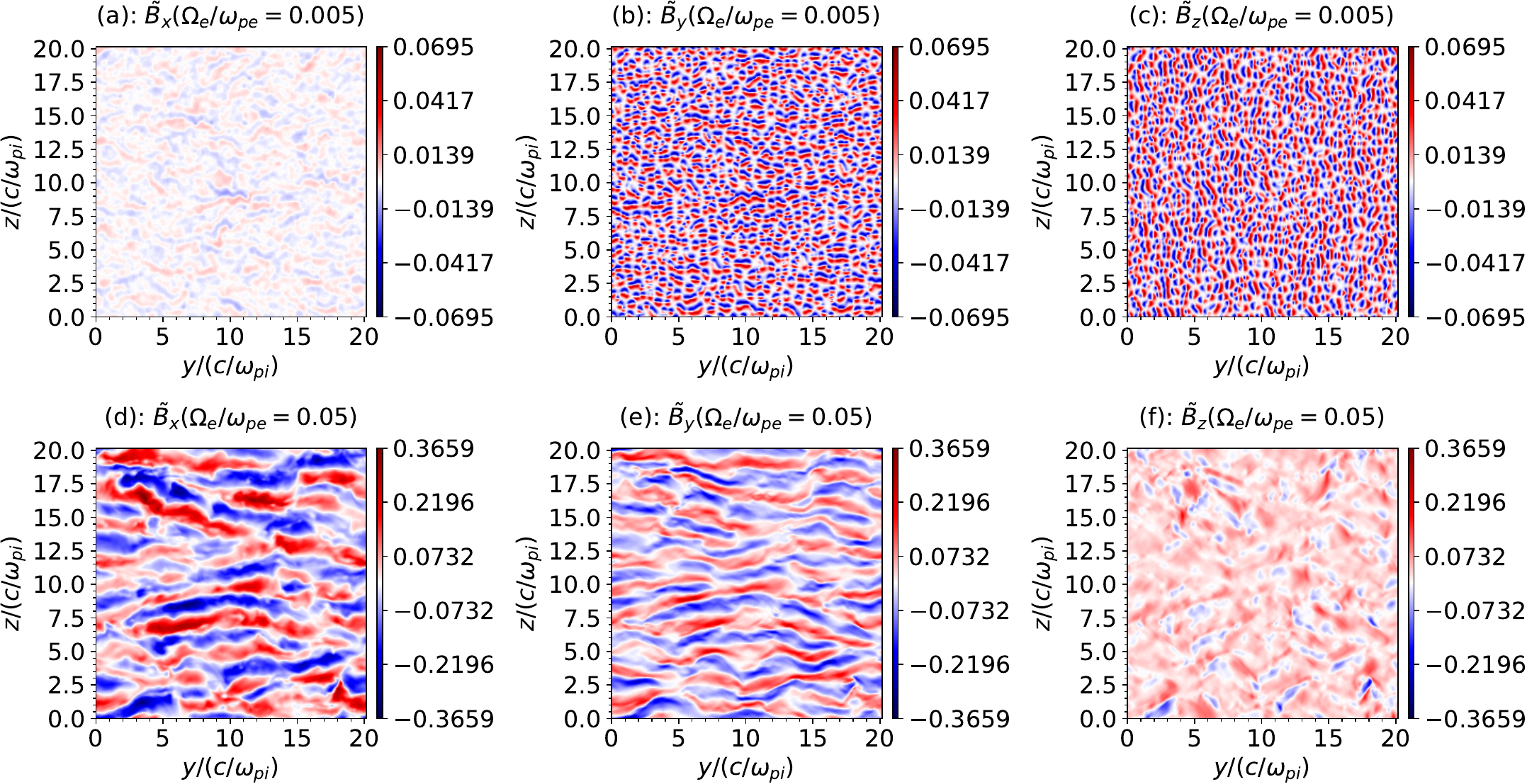}
\caption{\label{fig2} Snapshots of magnetic field at $\omega_{pi}t=250$. Top (a)-(c) and bottom (d)-(f) panels show $\tilde{B}_x, \tilde{B}_y, \tilde{B}_z$ for the unmagnetized ($\Omega_e/\omega_{pe}=0.005$) and the magnetized electron ($\Omega_e/\omega_{pe}=0.05$) cases, respectively.}
\end{figure}

First, we focus on the out-of-plane beam configuration.
Fig. \ref{fig2} shows the snapshots of the magnetic field at a fully nonlinear phase $\omega_{pi} t = 250$.
Panels (a)-(c) show the result of the unmagnetized electron case, which is consistent with the theoretical analysis (e.g., Fig. \ref{fig1} (b) and (d)) in the sense that the magnetic fluctuations are symmetric in the $y-z$ plane and the characteristic wavelength is on the order of the electron inertial length.
Panels (d)-(e) show the result of the magnetized electron case.
In this case, we can clearly see the dominance of parallel propagation (i.e., the wavenumber is primarily in the $z$ direction), which is again consistent with the theory (e.g., Fig. \ref{fig1} (c)).
Another key difference between the top and bottom rows is in the beam-aligned magnetic field $\tilde{B}_x$: It is almost absent for the unmagnetized electron case (c), while it is the largest component for the magnetized electron case (f). The latter is clearly not consistent with the theoretical prediction (e.g., Fig. \ref{fig1} (d)).

\begin{figure}[htbp]
\centering
\includegraphics[width=\linewidth]{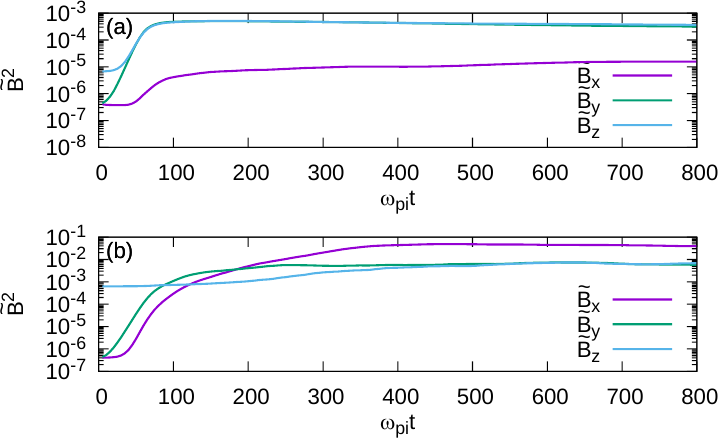}
\caption{\label{fig3} Time evolution of averaged magnetic field energy for the out-of-plane beam configuration for (a) unmagnetized electron, (b) magnetized electron cases. Violet, green, and cyan lines show the $x, y$, and $z$ components, respectively.}
\end{figure}

Fig. \ref{fig3} shows the time evolution of normalized magnetic field energy for each component.
Panel (a) shows the result of the unmagnetized electron case.
In this case, the weak background magnetic field in the $z$ direction hardly affects the time evolution, and the dominant components are always $\tilde{B}_y$ and $\tilde{B}_z$. This confirms that the background magnetic field, in this case, is sufficiently weak such that the electrons behave unmagnetized.
On the other hand, panel (b) for the magnetized electron case shows different behavior.
Initially, the dominant perturbation appears in the $y$ component shown with green and is consistent with the theory for the wavenumber vector parallel to the $z$ direction.
The $z$ component shown with cyan (parallel to the background magnetic field) does not show an appreciable change until the late nonlinear stage $(\omega_{pi}t \gtrsim 300)$.
The beam-aligned component $\tilde{B}^2_x$ in violet exhibits a somewhat different time evolution.
During the early stage $0\lesssim\omega_{pi}t\lesssim100$, $tilde{B}^2_x$ is roughly $\sim20\%$ of the primary component $\tilde{B}^2_y$, which is consistent with the linear eigenvector (Fig. \ref{fig1} (d)).
However, it continues to grow even after the saturation of $\tilde{B}_y^2$ at around $\omega_{pi}t\sim150$ and eventually becomes the dominant component.
We suggest that the growth of the beam-aligned component in the nonlinear stage may be understood in terms of a dynamo-like amplification mechanism.

Since we consider the regime where ions are unmagnetized, but electrons are magnetized, the electric field may reasonably be estimated by the electron frozen-in condition $\bm{E} = -\bm{v}_e \times \bm{B} / c$, where $\bm{v}_e$ is the electron flow velocity. 
Assuming that the electron flow is primarily in the $x$ direction (as they are dragged by the ion beam) and the inhomogeneity is in the $z$ direction as indicated by the simulation result, the beam-aligned component of the magnetic induction equation may be written as $\partial B_x/\partial t \approx \partial (v_{e,x} B_z) / \partial z$. We thus understand that a constant seed magnetic field $B_z$ will be dragged by an electron flow shear $\partial v_{e,x}/\partial z$ to generate the beam-aligned magnetic field component $B_x$. 
It is important to note that once the electron flow shear is established, this mechanism does not require instability driving. Since the electron flow itself is driven by ions, the magnetic field amplification will continue up to the point where the Lorentz force becomes strong enough to slow down the ion flow, which explains why the amplification of the beam-aligned component alone proceeds even after the saturation of $B_y$. 
We call this mechanism the electron magnetohydrodynamic (EMHD) dynamo because of its similarity with the kinematic MHD dynamo, but the magnetic field is frozen into the electron fluid rather than the bulk MHD fluid.
However, it is important to note that the simulation domain is limited to the shock transition region.
Therefore, the magnetic field amplification discussed here does not necessarily predict the far downstream (fully MHD scale) magnetic field structure. 

We may estimate the time scale of the EMHD dynamo amplification $T \equiv (d \tilde{B}_x /dt)^{-1}$ by substituting $\partial v_{e,x}/\partial z \sim V_{\mathrm{sh}}/\lambda_i$ (according to the simulation result) to obtain $\Omega_{i} T \sim N^{-1/2}$, which implies the scaling with $\Omega_{i}^{-1}$ in contrast to $\omega_{pi}^{-1}$ for Weibel instability. 
We have confirmed that the characteristic time scale of the initial Weibel instability phase is $\sim 100 \omega_{pi}^{-1}$, which is then followed by a slower growth up to $\sim \Omega_i^{-1}$ independent of the mass ratio (see Subsec. \ref{subsec:massratio}). 
Consequently, the EMHD dynamo converts a few percent of the initial ion flow kinetic energy to the magnetic field, indicating that $\delta B/B_0 \gg 1$ will naturally result within the transition region of high Mach number shocks. 
We further discuss the nonlinear magnetic field amplification, including comparisons with different simulation conditions in Sec. \ref{sec:saturation}.

\subsection{In-Plane Beam} \label{subsec:inplane}
\begin{figure}[htbp]
\centering
\includegraphics[width=\linewidth]{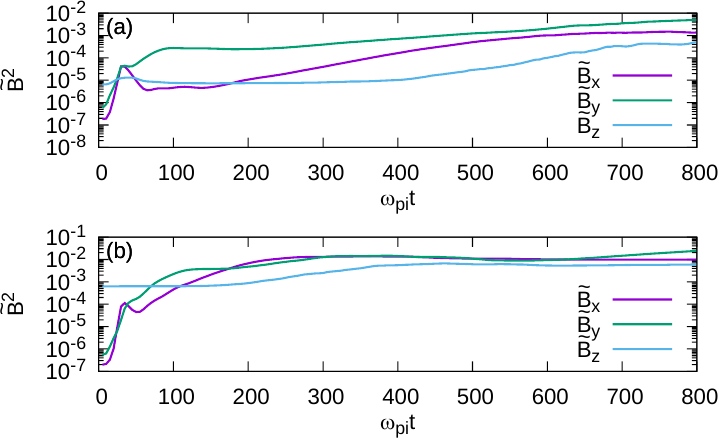}
\caption{\label{fig4} Time evolution of magnetic field energy for the in-plane beam configuration. The format is the same as Fig. \ref{fig3}}
\end{figure}

\begin{figure}[htbp]
\centering
\includegraphics[width=\linewidth]{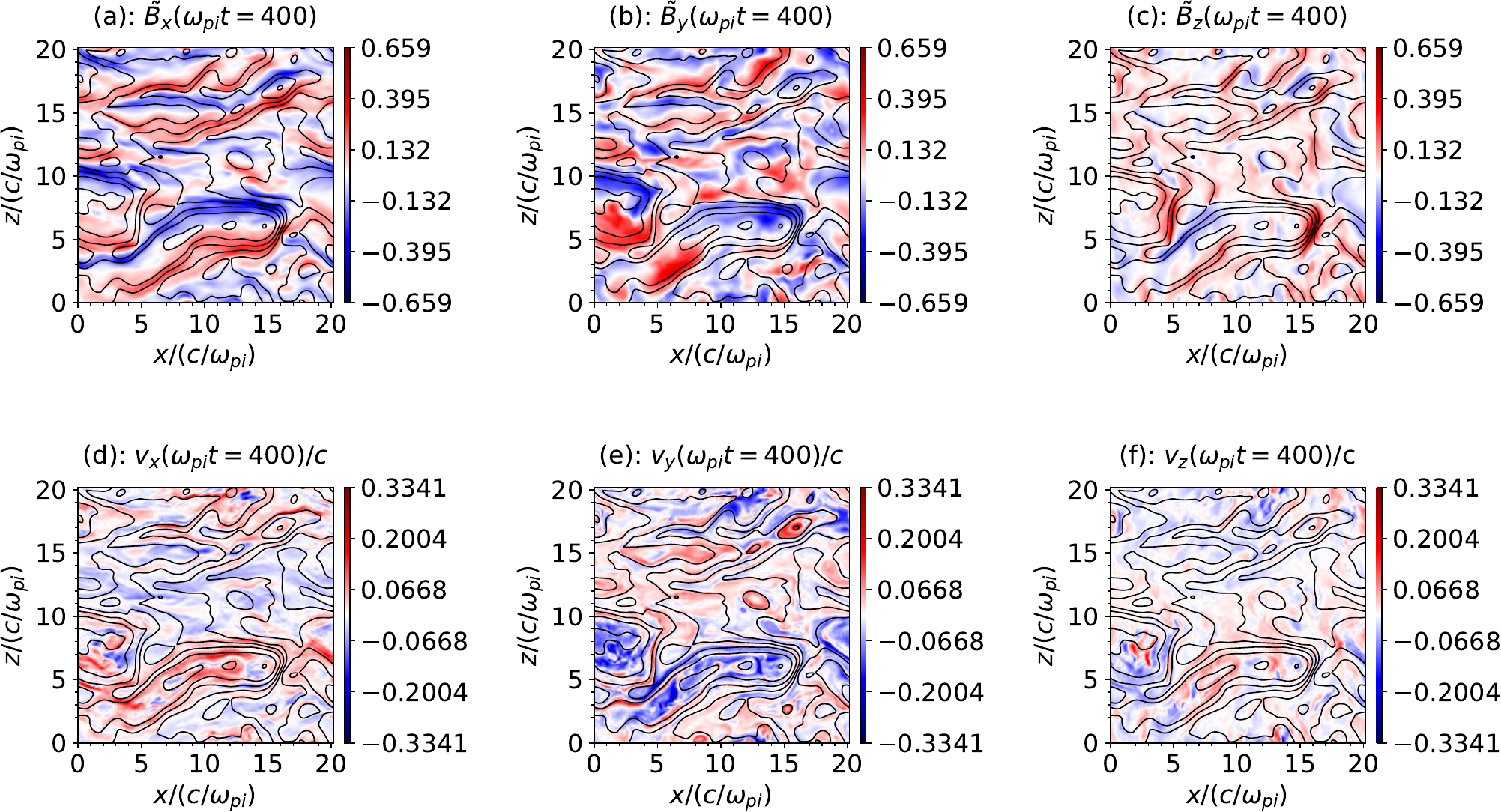}
\caption{\label{fig5} Snapshots taken at $\omega_{pi}t=400$. The top (a-c) panels show the magnetic field components, and the bottom (d-f) panels show the electron fluid velocity. The solid black lines indicate the in-plane magnetic field lines.}
\end{figure}

Although the current sheet structure associated with the amplification of the beam-aligned component is sustained in the out-of-plane beam configuration in the late nonlinear phase, this will not necessarily be realistic.
Here, we show the time evolution of the magnetic field for the in-plane beam case.
The time evolution of the magnetic energy for each component is shown in Fig. \ref{fig4} with the same format as Fig. \ref{fig3}.
Although the effect of electrostatic Buneman instability in the very early stage $\omega_{pi}t\lesssim40$ makes it difficult to identify the development of Weibel instability, the growth after the saturation of Buneman instability may be attributed to it.
For the unmagnetized electron case (panel (a)), the primary component $\tilde{B}^2_y$ is the dominant component after the saturation of Buneman instability $\omega_{pi}t \gtrsim 50$.
We note that the growth of $\tilde{B}^2_x$ after $\omega_{pi}t\sim200$ may also be understood as a result of a strong electron flow shear within the simulation plane.
However, we think this flow shear is an artifact of the 2D simulation because the cylindrical current structure (rather than the current sheet) will develop in fully 3D if the electrons behave unmagnetized \citep{Frederiksen_2004,Hededal_2004,Ruyer_2018,Takamoto_2018}.
The beam-aligned component $\tilde{B}^2_x$ for the magnetized electron case keeps growing well after $\tilde{B}^2_y$ reaches an apparent saturation at around $\omega_{pi}t\sim120$.
Magnetic reconnection is later triggered at around $\omega_{pi}t\sim200$ when $B_x\gg B_0$ is reached.
We can also confirm that the reconnected component $\tilde{B}^2_z$ starts to increase around this time as a result. 
In addition, we see characteristic patterns of magnetic reconnection in the electron velocity.
Fig. \ref{fig5} shows the snapshots taken at $\omega_{pi}t=400$.
We can see negative $B_z$ in panel (b), even though the background magnetic field $B_z$ is initially positive.
In panel (f), we see that the electron velocity pattern is coupled with the magnetic field lines. 
In particular, the magnetic islands are associated with the out-of-plane electron flow.

\begin{figure}[htbp]
\centering
\includegraphics[width=\linewidth]{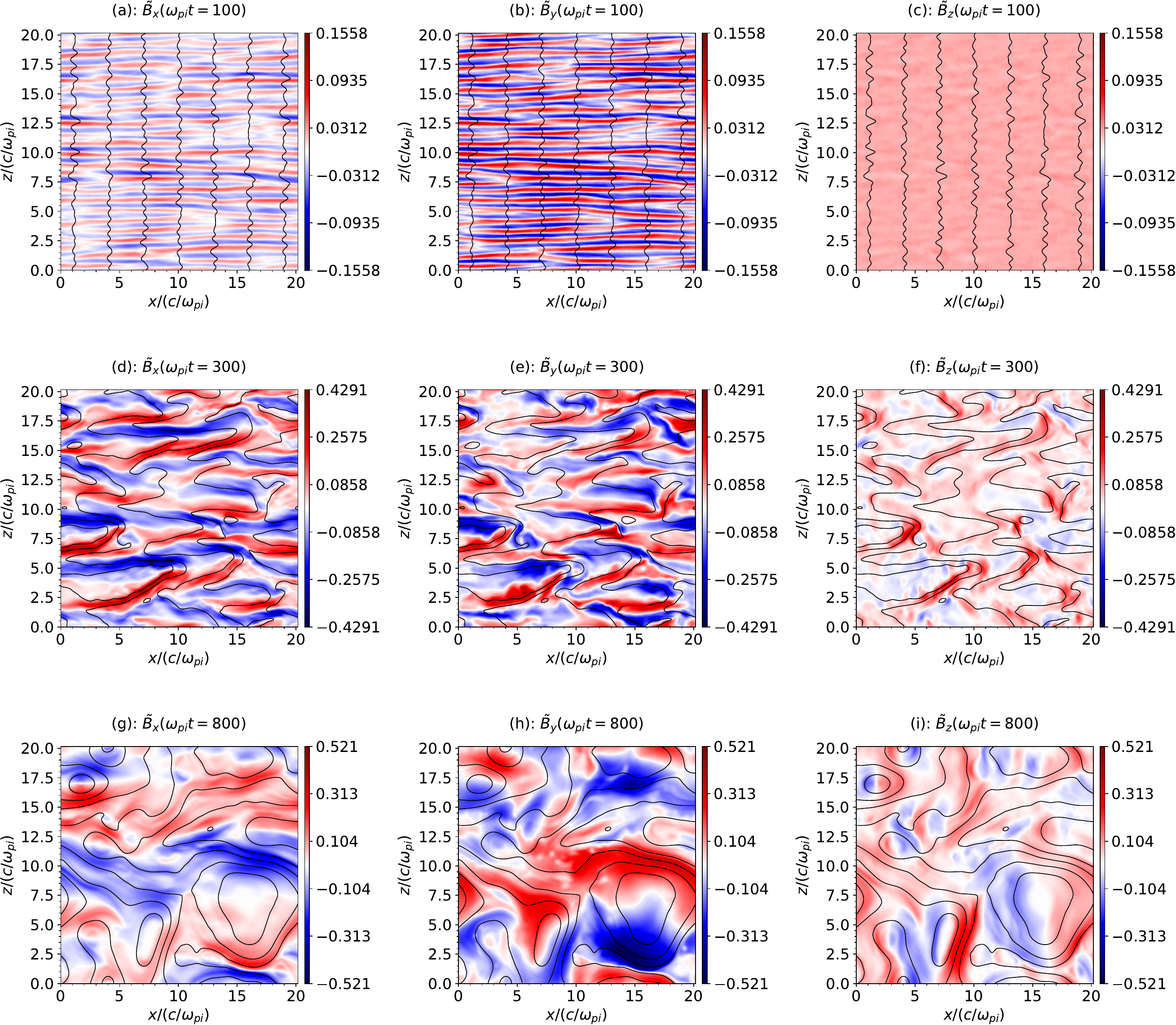}
\caption{\label{fig6} Snapshots of magnetic field for the in-plane beam configuration. Top (a)-(c), middle (d)-(f), and bottom (g)-(i) panels show snapshots taken at $\omega_{pi}t=100, 300, 800$, respectively. The solid black lines indicate the in-plane magnetic field lines.}
\end{figure}

The snapshots of the magnetic field for three characteristic times $\omega_{pi} t = 100, 300, 800$ for the in-plane beam configuration are shown in Fig. \ref{fig6}.
Panels (a)-(c) show the magnetic field in the early phase $\omega_{pi} t = 100$, which confirms that both the primary component $\tilde{B}_y$ and the beam-aligned component $\tilde{B}_x$ have substantial amplitudes.
Panels (d)-(f) at $\omega_{pi} t = 300$ visually demonstrate that the magnetic field lines are being dragged by the electron flow in the $x$ direction to generate the beam-aligned component by the EMHD dynamo mechanism.
At this time, we can see that some magnetic island structures are already created by magnetic reconnection at the current sheets.
As magnetic reconnection converts the beam-aligned component $\tilde{B}_x$ to $\tilde{B}_z$, fluctuations in $\tilde{B}_z$, particularly their negative excursions, may be recognized as the signature of magnetic reconnection.
Finally, Panels (g)-(i) at $\omega_{pi} t = 800$ show the magnetic field structure in the fully nonlinear stage.
As magnetic reconnection continuously proceeds, larger-scale magnetic islands have been generated.
We have confirmed that $\tilde{B}_z^2$ starts to increase before saturation and eventually reaches the same level as $\tilde{B}_x^2$. 
This indicates that magnetic reconnection plays an essential role in saturation. 
It is noted that the typical time scale of magnetic reconnection is given by the local ion gyroperiod determined by the amplified magnetic field strength. 
Therefore, we conclude that the system will evolve into the spontaneous magnetic reconnection phase within the dynamic time scale of the shock.

\subsection{Mass Ratio Dependence} \label{subsec:massratio}
\begin{figure}[htbp]
\centering
\includegraphics[width=\linewidth]{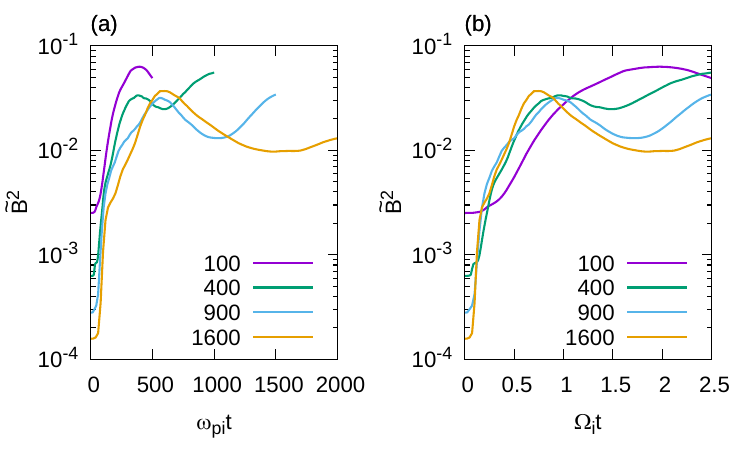}
\caption{\label{fig7} Temporal evolution of magnetic energy. Panel (a) uses $\omega_{pi}^{-1}$ as the unit of time and (b) uses $\Omega_{ci}^{-1}$. Violet, green, cyan, and orange lines correspond to $m_i/m_e=100,400,900,1600$, respectively.} 
\end{figure}

In Subsecs. \ref{subsec:inplane} \ref{subsec:outofplane}, we discussed the simulation result for a fixed mass ratio of $m_i/m_e=400$.
Here, we show the results of the in-plane beam simulations with various mass ratios ranging from 100 to 1600.
Fig. \ref{fig7} shows the evolution of magnetic field energy in the in-plane beam simulations.
$N=0.2,V_{\mathrm{sh}}=0.25c,\Omega_e/\omega_{pe}=0.05,v_{\mathrm{th},e}=0.1c,v_{\mathrm{th},i}=0.00125c$ is common for all the mass ratios.
We keep the shock velocity $V_{\mathrm{sh}}$ fixed so that the growth rate of the Weibel instability in units of $\omega_{pi}$ becomes the same for all mass ratios.
This choice leads to different Alfv\'{e}n Mach numbers; $M_{\mathrm{A}}=50,100,150,200$ for $m_i/m_e=100,400,900,1600$, respectively.
Note that all cases satisfy the magnetized electron condition $(\Gamma\ll\Omega_e)$.
Panel (a) uses $\omega_{pi}^{-1}$ as the unit of time.
We confirm that the characteristic time scale of the linear Weibel instability is $t\sim100\omega_{pi}^{-1}$ for all mass ratios.
Panel (b) shows the same results but uses $\Omega_i^{-1}$ as the unit of time.
Here we see that the characteristic time of the dynamo-like amplification is $T\sim\Omega_i^{-1}$ for $m_i/m_e=400,900,1600$.
The quantitatively different behavior of $m_i/m_e=100$ may imply that the scale of the magnetized electron instability $5\lesssim k\lambda_i\lesssim10$ is not well separated from the electron inertial scale $k\lambda_i\sim\sqrt{m_i/m_e}=10$.
The dynamo-like amplification of the magnetic field completes within $\Omega_i^{-1}$ calculated using the initial background magnetic field, even with a realistic mass ratio.
This implies that nonlinear amplification can occur in the shock dynamical time scale.
The three mass ratios $400,900,1600$ also share a similar amount of magnetic-field amplification ($\sim 4\%$ of the kinetic energy).

\section{Nonlinear Saturation} \label{sec:saturation}
In this section, we discuss the nonlinear saturation of the magnetic field.
We used periodic boundary conditions and initial conditions uniform in space.
This type of simulation was used in previous studies to investigate micro-instabilities in magnetized shock transition regions. 
It has been shown that the model reasonably reproduces the waves and instabilities seen in the full shock transition region \citep{Matsukiyo_2003,Scholer_2003,Shimada_2000,Shimada_2004}.
However, it is also known that the late nonlinear evolution of the magnetic field could be more complicated in more realistic setups.
Here, we shall compare our results to those with different setups and clarify the role of the background magnetic field.

First, let us discuss the unmagnetized case.
In our periodic boundary simulation, the Weibel instability resulted in $\tilde{B}^2\sim 0.001$ (Fig. \ref{fig3} (a), \ref{fig4} (a)).
We have also run simulations with different $n_r/n_0$ and electrons initially drifting in the same velocity as the ions (instead of thermal background).
$\tilde{B}^2$ was not sensitive to these conditions.
These results are consistent with the estimate by trapping (Fig. \ref{fig1} (d)) and previous periodic boundary simulations, e.g.,\citet{Ruyer2015}.
Note that the saturation level in the unmagnetized case strongly depends on the mass ratio.
Considering the different mass ratios in simulations, the saturation level in open boundary shock simulations is also consistent with our result, at least for the early stages of shock formation \citep{Kato_2008,Ruyer2016}.
The dynamics of an unmagnetized shock after saturation of Weibel and the conditions for ignoring the background magnetic field need further investigation, which we will report in a future paper.

Now, we move on to the magnetized case.
Our periodic simulation showed $\tilde{B}^2\sim0.05$ (Fig. \ref{fig3} (b), \ref{fig4} (b)).
This is notably larger than the estimate by trapping (Fig. \ref{fig1} (d)).
We could understand the extra amplification by the EMHD-like process.
In Sec. \ref{sec:simulation}, we discussed that the saturation of the magnetized system is defined by the energy balance between the beam kinetic energy and magnetic energy (out-of-plane beam case) or dissipation by spontaneous magnetic reconnection (in-plane beam case).
Conclusive discussions of the late saturation need fully 3D simulations, which will be discussed in a forthcoming paper.
We note that 2D (magnetized) perpendicular shock simulations showed slightly larger values up to $\tilde{B}^2\sim 0.1$ \citep{Bohdan_2021}, which may be attributed to, for instance, different mass ratios or continuous energy input.

From these inspections, we conclude that electron magnetization is clearly responsible for the more efficient magnetic field amplification in magnetized shocks.
Furthermore, the qualitatively different magnetic field structures, such as the dominance of the beam-aligned field and spontaneous magnetic reconnection onset, clearly distinguish the weakly magnetized shocks from completely unmagnetized shocks.

\section{Discussions and Conclusions} \label{sec:conclusion}
In this paper, we have investigated the role of a finite background magnetic field in the linear and nonlinear evolution of ion-beam Weibel instability.
The linear analysis found that, if the electrons behave magnetized, the finite background magnetic field shifts the dominant wavelength from the electron inertial scale to the intermediate scale $5 \lesssim k\lambda_i \lesssim 10$. 
Furthermore, it breaks the symmetry with respect to the beam direction. 
Indeed, nonlinear PIC simulations demonstrate that the waves propagating along the background magnetic field grow preferentially into large amplitudes. 
We have found that the beam-aligned magnetic field component continues to grow via the EMHD dynamo amplification of a seed in-plane magnetic field component even after the primary component of Weibel instability saturates, resulting in the formation of coherent current sheets. 
The amplified magnetic field energy eventually reaches a few percent of the initial free energy in the system, and if the beam-aligned component is contained within the 2D simulation plane, magnetic reconnection is spontaneously triggered (Subsec. \ref{subsec:inplane}). 
The nonlinear evolution is fundamentally different from the unmagnetized Weibel instability, in which large-scale filamentary currents are generated and eventually disrupted via the kink instability \citep{Ruyer_2018}.  
A preliminary 3D PIC simulation result for the magnetized electron case has also confirmed that current sheets (rather than filaments) form in the nonlinear phase.

We have found previously that the condition for a magnetized Weibel-dominated shock is roughly given by $M_{\mathrm{A}} \gtrsim 32 (N/0.2)^{-1/2}$ \citep{Amano2022}.
The dependence on electron magnetization indicates that the Weibel-dominated shock may be further classified into two types.
Using the analytic form of the maximum growth rate $\Gamma_{\mathrm{max}}/\omega_{pi} \sim N^{1/2} V_{\mathrm{sh}}/c$, we find that the transition between the magnetized and unmagnetized electron regimes occurs at $$M_{\mathrm{A}} \sim N^{-1/2} m_i/m_e\sim 4100 (N/0.2)^{-1/2} (m_i/m_e / 1836).$$ Typical Alfv\'{e}n speeds in the interstellar medium $V_{\mathrm{A}}/c \sim 10^{-4}{\rm -}10^{-3}$ indicate that electrons in a strong astrophysical shock will behave magnetized unless the shock speed becomes relativistic. Therefore, the spontaneous magnetic reconnection scenario first found by \citet{Matsumoto_2015} is indeed applicable to non-relativistic SNR shocks with 
$32 (N/0.2)^{-1/2} \lesssim M_{\mathrm{A}} \lesssim 4100 (N/0.2)^{-1/2} (m_i/m_e / 1836)$.
On the other hand, external shocks of relativistic jets in AGN and GRBs propagating in weakly magnetized media will be dominated by classical unmagnetized Weibel turbulence without involving magnetic reconnection.

Future investigation will explore the nonlinear evolution in fully 3D to see how the different electron responses affect the overall evolution of the system and the efficiency of particle acceleration. 
While our finding is difficult to prove with astrophysical observations alone, laboratory laser experiments may provide the possibility \citep{Fiuza_2020}. 
It is important to point out that since the transition between the two regimes is dependent on the ion-to-electron mass ratio, the ion composition in the laboratory plasma, as well as the artificial mass ratio often adopted in PIC simulations, must be carefully taken into account. 
With the caveat, the laboratory experiment combined with fully kinetic simulations will be a promising tool to investigate particle acceleration at high Mach number astrophysical shocks.

\appendix
\section{Polarization} \label{subsec:polarization}
Here, we discuss the polarization of the Weibel magnetic field in the linear stage.
First, we start with the unmagnetized case.
The normalized eigenvector is given by $\tilde{\bm{E}}=(1,0,0)^{\mathsf{T}}$  (see Eq. (\ref{dispersiontensor})).
Note that Faraday's law (or $\tilde{\bm{B}}=c\bm{k}\times\tilde{\bm{E}}/\omega$) indicates that the primary component of the magnetic field is perpendicular to both the beam and the wavenumber.
This result is consistent with the dominance of the beam perpendicular magnetic field ($B_x, B_y$ for the out-of-plane beam case (Fig. \ref{fig2})).

\begin{figure}[htbp]
\centering
\includegraphics[width=0.5\linewidth]{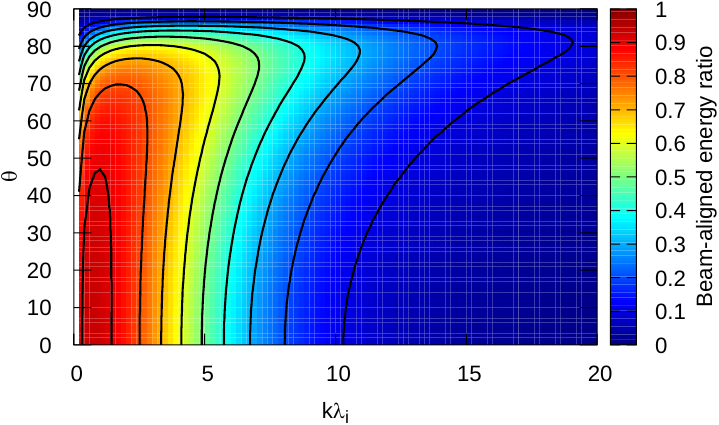}
\caption{\label{figA1} Energy ratio of beam-aligned component during the linear stage $|\tilde{B}_x|^2/(|\tilde{B}_x|^2+|\tilde{B}_y|^2+|\tilde{B}_z|^2)$} plotted in the $k-\theta$ plane.
\end{figure}

When there is a finite background magnetic field, the polarization deviates from the linear polarization predicted for the unmagnetized plasmas.
The ions are unaffected by the background magnetic field, i.e., $\Gamma_{\mathrm{max}}/\Omega_i\ll 1$ is still satisfied.
On the other hand, the magnetized electrons generate beam-perpendicular current due to their $\delta\bm{E}\times\bm{B}_0$ drift, which produces a beam-alined magnetic field component.
In Fig. \ref{figA1}, we show the energy ratio of the beam-aligned magnetic field defined by $$\frac{|\tilde{B}_x|^2}{|\tilde{B}_x|^2+|\tilde{B}_y|^2+|\tilde{B}_z|^2}=\frac{|\cos{\theta}\tilde{E}_y-\sin{\theta}\tilde{E}_z|^2}{|\tilde{E}_x|^2+|\cos{\theta}\tilde{E}_y-\sin{\theta}\tilde{E}_z|^2}.$$
The saturation level estimate for the beam-aligned component, the red dotted line in Fig. \ref{fig1} (d), is obtained by Eq. (\ref{eq:saturation_level}) multiplied by this ratio.
The short wavelength modes are hardly affected by the background magnetic field, and the beam-aligned component is almost negligible for $10<k\lambda_i$.
For wavenumbers $5\lesssim k\lambda_i\lesssim10$, in which the saturation level is largest, there is a finite beam-aligned component of $10\%$ to $50\%$ which is consistent with the linear stages of simulation (e.g., Fig. \ref{fig3} (b)).
Note that this discussion is limited to the linear stage.
We have shown that the beam-aligned component could dominate after the dynamo-like amplification.
The beam-aligned component is dominant for wavelength longer than the ion inertial scale.
However, we do not put emphasis on these modes since theory predicts these modes have low saturation levels \citep{Lyubarsky2006}.
We also confirmed that these small wavenumber modes had small amplitude in the simulations.

\begin{acknowledgments}
This work was supported by JSPS KAKENHI Grants No. 17H06140, 22K03697, and 	22J21443. 
T.~J. is grateful to The International Graduate Program for Excellence in Earth-Space Science (IGPEES), University of Tokyo. 
T.~A. thanks for the support from the International Space Science Institute (ISSI) in Bern through ISSI International Team Project \#520 ({\it Energy Partition across Collisionless Shocks}). 
This research was also supported by MEXT as “Program for Promoting Researches on the Supercomputer Fugaku” (Toward a unified view of the universe: from large scale structures to planets, JPMXP1020200109) and JICFuS.
\end{acknowledgments}

\bibliography{reference}

\end{document}